\documentclass[preprint, showpacs, prb]{revtex4-1}
\usepackage{amsmath}
\usepackage{bm}
\usepackage{hyperref}
\usepackage{url}
\usepackage{subfigure}
\usepackage{graphicx}
\begin{document}
\title{Continuous phase transition from N\'eel state to Z$_2$
  spin-liquid state on a square lattice}
\date{\today}
\author{Yang Qi}
\affiliation{Institute for Advanced Study, Tsinghua University, Beijing 100084, China}
\author{Zheng-Cheng Gu}\thanks{Present Address: Perimeter Institute
  for Theoretical Physics, Waterloo, Ontario, N2L 2Y5, Canada}
\affiliation{Institute for Quantum Information and Matter, California
  Institute of Technology, Pasadena, CA 91125, USA}

\begin{abstract}
  Recent numerical studies of the $J_1$-$J_2$ model on a square
  lattice suggest a possible continuous phase transition between the
  N\'eel state and a gapped spin-liquid state with Z$_2$ topological
  order.  We show that such a phase transition can be realized through
  two steps: First bring the N\'eel state to the U(1) deconfined
  quantum critical point, which has been studied in the context of
  N\'eel -- valence bond solid (VBS) state phase transition. Then
  condense the spinon pair -- skyrmion/antiskyrmion bound state, which
  carries both gauge charge and flux of the U(1) gauge field emerging
  at the deconfined quantum critical point. We also propose a
  Schwinger boson projective wave function to realize such a Z$_2$ spin
  liquid state and find that it has a relatively low variational
  energy($-0.4893J_1$/site) for the $J_1$-$J_2$ model at
  $J_2=0.5J_1$. The spin liquid state we obtain breaks the fourfold
  rotational symmetry of the square lattice and therefore is a nematic
  spin liquid state. This direct continuous phase transition from
  the N\'eel state to a spin liquid state may be realized in the $J_1$-$J_2$
  model, or the anisotropic $J_{1x}$-$J_{1y}$-$J_2$ model.
\end{abstract}

\maketitle

A spin liquid state has been searched for both theoretically and
experimentally for decades, especially for the purpose of
understanding the novel mechanism of high-$T_c$
cuprates\cite{LNW2006Review}.  One of the most interesting and
relevant models is the $J_1$-$J_2$ spin-$1/2$ antiferromagnetic
Heisenberg model on a square lattice, since the frustration induced by
the $J_2$ term in the $J_1$-$J_2$ model might mimic the frustration
induced by the hopping term in the $t$-$J$ model, which has been
believed to be the low-energy effective model of high-$T_c$
cuprates\cite{Zhang1988}. According to Anderson's resonating valence
bond (RVB) scenario\cite{Anderson1973}, the potential spin liquid
state in the $J_1$-$J_2$ model might be the most important low-energy
metastable state of cuprates and the superconducting ground state will
be naturally developed upon doping\cite{Anderson1987}. On the other
hand, the $J_1$-$J_2$ model can be realized in many frustrated
magnets\cite{Dai2012,Melzi2001}; thus investigating the phase diagram
of such a simple model would be of great importance by
itself. Previous theoretical studies using the mean-field theory have
found a possible Z$_2$ spin liquid phase in the $J_1$-$J_2$
model\cite{rs1, Sachdev1991,Flint2009}. Very recently, a spin liquid
ground state has been observed in the maximal frustrated region($J_2
\sim 0.5J_1$) by numerical studies\cite{Jiang2012,Wang2011}. The
discovered spin liquid ground state has gaps in both spin singlet and
triplet channels, and a universal constant $\gamma\simeq\ln2$ in the
entanglement entropy. These signatures indicate a gapped spin liquid
with Z$_2$ topological order. Moreover, the numerical studies also
show evidences for a continuous phase transition between the N\'eel
state with antiferromagnetic ordering at the wave vector $(\pi, \pi)$,
and the (possible) Z$_2$ spin liquid state.

Studies of quantum phase transitions between quantum spin liquid
phases and adjacent phases are important for the understanding of the
spin liquid states, as they provide vital information on the effective
field theory description of the spin liquid and also predict universal
behaviors that can be compared with experimental and numerical
results.  However, in the past there has been no theory that can
describe a continuous phase transition between the N\'eel state and a
Z$_2$ spin liquid state in a model with the SU(2) spin rotational
symmetry. Particularly, the theory of deconfined quantum criticality
indicates that killing the antiferromagnetic order in the N\'eel state
does not result in a symmetric paramagnetic state but a valence bond
solid (VBS) state\cite{senthil2004, senthil2004a}. On the other hand,
starting from a bosonic Z$_2$ spin liquid state, one can bring it to
an antiferromagnetic state through a continuous phase transition by
condensing the spinon excitations, but the resulting antiferromagnetic
state has a noncollinear order\cite{Sachdev2008, xucoming}, rather
than the collinear order that the N\'eel state has. It is not until
the work by \citet{Moon2012} that a continuous phase transition
between a Z$_2$ spion liquid and a collinear antiferromagnetic state
is proposed. In their theory they show that condensing bound states of
spinon and vison excitations in the Z$_2$ spin liquid state leads to a
continuous phase transition to a collinear antiferromagnetic
state. However, their study is based on a field theory analysis and it
is not clear what kind of specific SU(2) symmetric lattice model can
support such a field theory.

\begin{figure}[htbp]
  \centering
  \includegraphics[width=12cm]{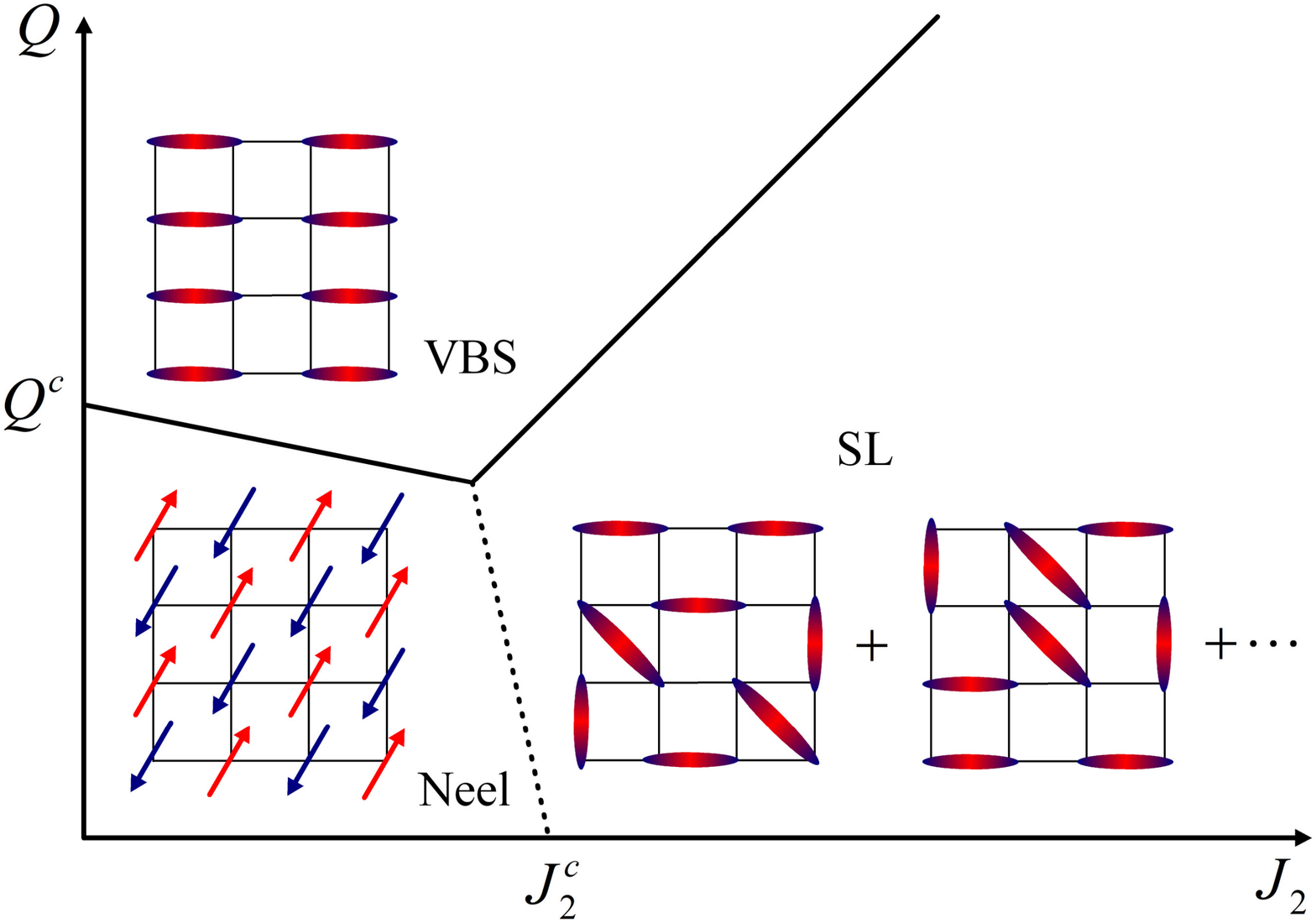}
  \caption{Conjectured phase diagram of the $J_1$-$J_2$-$Q$ model. In
    the phase diagram we set $J_1=1$ and vary the other two
    frustration terms. At the origin $J_2=Q=0$ the model is in the
    N\'eel state. Along the $x$ axis $Q=0$ and the model reduces to
    the $J_1$-$J_2$ model, which has a continuous phase transition
    between N\'eel and Z$_2$ spin liquid
    states\cite{Jiang2012,Wang2011}. Along the $y$ axis $J_2=0$ and
    the model reduces to the $J$-$Q$ model, which has a continuous
    phase transition between N\'eel and VBS
    order\cite{Sandvik2010a}. The solid lines show phase boundaries
    described by the deconfined quantum
    criticality\cite{senthil2004,senthil2004a}, and the dashed line
    shows the phase boundary that is the subject of this study, in
    which we propose that it can also be described by the deconfined
    quantum criticality.}
  \label{fig:cpd}
\end{figure}

In this work, we study the continuous phase transition between the
N\'eel and the Z$_2$ spin liquid state on square lattice starting from
the N\'eel state. We propose that the critical point of this phase
transition is described by the same deconfined quantum critical theory
that is also applicable to the critical point between the N\'eel and
the VBS order. As a motivation, we consider a $J_1$-$J_2$-$Q$ model
that contains both next-nearest-neighbor interaction temrs and
plaquette ring-exchange terms with coefficient $Q$. When $Q=0$, this
model is reduced to the $J_1$-$J_2$ model which has a phase transition
from the N\'eel to the Z$_2$ spin liquid phase. When $J_2=0$, the
$J$-$Q$ model has been studied by the quantum Monte Carlo
method\cite{Sandvik2010a} and it realizes the continuous phase
transition from N\'eel to VBS phase described by the deconfined
quantum critical theory. Based on these two limits we can conjecture a
possible phase diagram of the $J_1$-$J_2$-$Q$ model, as illustrated in
Fig.~\ref{fig:cpd}, assuming that there are no other phases between
the two limits and all phase transitions are of second order. In the
phase diagram the phase boundaries between the N\'eel and the VBS
state and between the VBS and the Z$_2$ spin liquid
state\cite{senthil2004a} are both described by the theory of
deconfined quantum criticality. As these two phase boundaries are
connected to the phase boundary separating the N\'eel and the spin
liquid state, it is likely that the latter is also described by the
same deconfined quantum critical point. We note that a numerical study
on the $J_1$-$J_2$-$J_3$ model\cite{Mambrini2006} gives evidence for a
similar phase diagram that contains the N\'eel phase, a plaquette VBS
phase and possiblely a Z$_2$ spin liquid phase.

Moreover, we propose that the Z$_2$ spin liquid state is obtained from
the deconfined quantum critical point by condensing the spinon
pair--skyrmion/antiskyrmion bound state. In the theory of deconfined
quantum criticality, the effective theory of the critical point is a
CP(1) model that contains a spin-$\frac12$ spinon field coupling to an
emergent U(1) gauge field. Starting from this deconfined quantum
critical point, one can gap out the spin excitations by proliferating
topological defects known as the skyrmion and drive the system into
the VBS state. On the other hand, one can also obtain a Z$_2$ spin
liquid state by condensing a pair of spinon excitations, which acts as
a Higgs field carrying gauge charge $2e$ of the emergent U(1) gauge
field\cite{Sachdev2008}. To achieve these two goals simultaneously, we
propose a scenario where a Z$_2$ spin liquid state can be obtained
from the deconfined quantum critical point by condensing the the
spinon pair--skyrmion/antiskyrmion bound state.

One interesting feature of the Z$_2$ spin liquid state obtained in our
study is that it breaks the four-fold rotational symmetry of the
square lattice, or in other words it is a nematic spin liquid. This
result is obtained by a symmetry analysis in Section~\ref{sec:bs}, and
it is consistent with previous mean field studies\cite{rs1,
  Sachdev1991, Flint2009}. Therefore we predict that on the square
lattice if a gapped Z$_2$ spin liquid state is separated from the
N\'eel state by a continuous phase transition, the spin liquid state
should be nematic. We would like to emphasize that our theoretical
study is generic and is not tied to any particular model Hamiltonian,
though numerical evidences strongly suggest that it is very likely to
be realized in the $J_1$-$J_2$ model and the anisotropic
$J_{1x}$-$J_{1y}$-$J_2$ model. A detailed discussion will be presented
in Section~\ref{sec:vmc} and \ref{sec:conclusion}.

The rest of the paper is organized as the following: In
Section~\ref{sec:bs} we discuss the scenario of a continuous phase
transition from the N\'eel state to the Z$_2$ spin liquid state
through bound-state condensation. We first briefly review the spinon
and skyrmion/antiskyrmion excitations at the deconfined quantum
critical point, and then discuss the scenario of obtaining a Z$_2$
spin liquid state from the deconfined quantum critical point by
condensing the bound state of a spinon pair and a
skyrmion/antiskyrmion. By studying the projective symmetry group (PSG)
properties of the bound-state operators we identify the symmetry of
the Z$_2$ spin liquid state. It turns out that the obtained Z$_2$ spin liquid
state preserves all lattice symmetries except the fourfold
rotational symmetry of the square lattice, and it is therefore a
nematic spin liquid state.

In Sec.~\ref{sec:eff} we study the phase transition to the Z$_2$
spin liquid phase and the excitations in the spin liquid phase. We
argue that a spin liquid phase can be obtained from the U(1)
deconfined quantum critical point by proliferating spinon
pair--skyrmion/antiskyrmion bound states. We also find two types of
low-energy excitations in the Z$_2$ spin liquid state: spinons
carrying spin-$\frac12$ and visons that are vortex excitations of
the bound-state condensate. In our theory both the spinon gap and
vison gap close at the critical point, which is consistent with the
numerical studies\cite{Jiang2012,Wang2011}.

In Sec.~\ref{sec:sbmf} we construct a projective wave function for the
Z$_2$ spin liquid state that we obtain by condensing the bound-state
operator. The Schwinger boson projective wave function is a
well-established way to describe the N\'eel state and adjacent spin
liquid states\cite{Auerbach1988,arovas1988}, and it has been used to
study the $J_1$-$J_2$ model on a square lattice\cite{Flint2009,
  Li2012}. Near the N\'eel state there are several different Schwinger
boson projective wave functions describing Z$_2$ spin liquid states
with different topological orders, and they can be classified using
their PSG\cite{wenpsg, wang2007}. By matching the PSG of the
projective wave function to the PSG of the bound-state operator in the
effective theory, we are able to identify the particular Schwinger
boson projective wave function that represents the Z$_2$ spin liquid
state to which the N\'eel state can be connected through a continuous
phase transition.

In Sec.~\ref{sec:vmc} we study the Schwinger boson projective wave
function using the variational Monte Carlo method. Our calculation is
based on the nonorthogonal valence bond basis\cite{Liang1988}, where
the sign problem is manageable if the state is close to the U(1)
deconfined quantum critical point. We show that this bosonic spin
liquid state has a relatively low ground-state energy, and it can be
stabilized by an anisotropy in the nearest-neighbor Heisenberg coupling
$J_{1x}\neq J_{1y}$.

\section{Bound state of spinon-pair and skyrmion.}
\label{sec:bs}

The starting point of our work is the theory of the deconfined quantum
criticality introduced by Senthil {\it et al} in
Ref. \onlinecite{senthil2004, senthil2004a}. Its main result is that
the critical point between the N\'eel state and the VBS state is described by
a non-compact CP(1) model that contains deconfined spin-$\frac12$
spinon fields coupled to an emergent non-compact U(1) gauge field. The
CP(1) model has the following Lagrangian,
\begin{equation}
  \label{eq:cp1}
  \mathcal L=\frac1g\sum_{\alpha=\uparrow\downarrow}
  |(\partial_\mu-ia_\mu)z_\alpha|^2,
\end{equation}
where $z_\alpha$ is a bosonic spinon field carrying spin-$\frac12$ and
it is related to the N\'eel order parameter $\bm n\sim(-1)^i\bm S_i$
in the following way,
\begin{equation}
  \label{eq:zn}
  \bm n=z_\alpha^\ast\bm\sigma_{\alpha\beta}z_\beta.
\end{equation}
The gauge field $a_\mu$ in Eq.~\eqref{eq:cp1} is an emergent U(1)
gauge field.

Another important part in the deconfined quantum criticality is the
topological excitation in the N\'eel state, called the
skyrmion. Skyrmion excitations are characterized by the skyrmion
number $Q$, a topological invariant of the spatial configuration of
the N\'eel order parameter $\bm n$, defined as the following,
\begin{equation}
  \label{eq:Qdef}
  Q=\frac1{4\pi}\int d^2x\bm n\cdot\partial_x\bm n\times\partial_y\bm n.
\end{equation}
The physical meaning of $Q$ is the total number of skyrmion
excitations, and it is conserved for smooth space-time configurations
of $\bm n$. However, in a lattice model, singular configurations of
$\bm n$ with tunneling events between configurations with different
skyrmion numbers are allowed. Therefore, in an effective theory, one
needs to add by hand skyrmion creation and annihilation events. In the
CP(1) model, skyrmion excitations are related to the gauge flux of
$a_\mu$ because of the following relation,
\begin{equation}
  \label{eq:2pQa}
  2\pi Q = \int d^2x(\partial_xa_y-\partial_ya_x).
\end{equation}
Hence we can relate skyrmion excitations to $2\pi$ flux quanta of the
$a_\mu$ gauge field. The existence of skyrmion tunneling events is
then equivalent to the existence of monopole events in the space-time
configuration of the gauge field, or to the fact that the gauge field is
compact.

The key result of the deconfined quantum criticality theory is that
the skyrmion creation and annihilation events are irrelevant at the
critical point, or in other words, the emergent U(1) gauge field is
non-compact. The reason behind this is the non-trivial Berry phase
associated with the skyrmion tunneling events\cite{haldane1988a},
which takes four different values on four sublattices of the dual
lattice. Because of this spatially dependent Berry phase,
contributions of skyrmion tunneling events cancel each other unless
the skyrmion number is changed by a multiple of four. As a result,
skyrmion tunneling events become irrelevant at the critical
point. Another consequence of this spatially dependent Berry phase is
that the proliferation of skyrmion excitations leads to the breaking
of lattice translational and rotational symmetry, and brings the
system to the VBS state. This effect can be understood by considering
the symmetry transformation of the skyrmion creation operator. The
Berry phase associated to skyrmion tunneling events results in a
non-trivial phase acquired by the skyrmion operator $v$ after lattice
symmetry transformations\cite{senthil2004a}, as summarized in
Table~\ref{tab:psg-field}. As a result, $v$ can be related to the
following linear combination of the order parameters of columnar VBS
states since they have the same symmetry
transformations\cite{senthil2004a}
\begin{equation}
  \label{eq:vvxy}
  v = e^{i\frac\pi4}(v_x+iv_y),
\end{equation}
where $v_x$ and $v_y$ denote the order parameters for columnar VBS
states in the $x$ and $y$ direction respectively. Hence the condensation of
$v$ leads to lattice symmetry breaking and therefore a VBS order.

Next, we discuss the scenario of obtaining a Z$_2$ spin liquid state
from the deconfined quantum critical point through condensing a bound
state of a skyrmion/antiskyrmion and a spinon pair. Starting from the
deconfined quantum critical point, which has an emergent U(1) gauge
field, a generic way of obtaining a Z$_2$ state is to condense a Higgs
field that carries gauge charge $2e$\cite{Sachdev2008}. On the other
hand, in order to kill the N\'eel order, we will need to condense the
skyrmion field. Consequently, we consider condensing a bound state of
these two excitations, which can be expressed as a product of the two
operators.

In the CP(1) model, a natural candidate of a charge-$2e$ Higgs field
is a pair of spinons. Since we are trying to get a spin liquid state,
the Higgs field must be a spin singlet. Hence the field must contain
at least one spatial derivative\cite{Sachdev2008}. The possible forms
at the lowest order are,
\begin{equation}
  \label{eq:umu}
  u_i = \epsilon_{\alpha\beta}z_\alpha\partial_i z_\beta, i=x,y.
\end{equation}

Now we can write a bound-state operator as a product of
skyrmion/antiskyrmion and spinon pair operators in Eq.~\eqref{eq:vvxy}
and~\eqref{eq:umu}. Actually there are more than one way to combine a
skyrmion/antiskyrmion and a pair of spinons, as both the
skyrmion/antiskyrmion and spinon pair fields have different
components. This can be resolved by analyzing how the bound-state
operator transforms under lattice symmetry operations. Since the Z$_2$
spin liquid state is obtained by condensing the bound-state operator,
its symmetry transformations determine the symmetry of the spin liquid
state. In order to obtain a spin liquid state with all lattice
symmetries, we search for a bound-state operator that is invariant
under lattice symmetry transformations.

One complication in the symmetry analysis of the bound-state operator
is that because of the gauge charge it carries, it can carry a
projective representation of the symmetry group\cite{wenpsg}, and
therefore does not need to be in the trivial representation to be
invariant under a symmetry operation. Particularly, the skyrmion
operator acquires a non-trivial phase under the translation and
condensing the skyrmion breaks the translational
symmetry\cite{senthil2004a}. However, although the bound-state
operator acquires the same phase under translation, such a phase can
be canceled by a U(1) gauge transformation and the spin liquid state
can still be translational invariant. Consequently, by condensing a
bound state instead of the skyrmion alone, the translational symmetry
is restored and a spin liquid state instead of the VBS state is
obtained. As an example, consider the $v_x$ component of the skyrmion
operator $v$, as defined in Eq.~\eqref{eq:vvxy}, which acquires a
minus sign upon translation in the $x$ direction,
\begin{equation}
  \label{eq:vxTx}
  T_x: v_x\rightarrow -v_x,
\end{equation}
and such symmetry transformation results in the translational symmetry
breaking of the VBS states obtained by condensing $v_x$. On the other
hand, the product of $u_i$ and $v_x$ carries gauge charge $2e$, and
the minus sign that appears in Eq.~\eqref{eq:vxTx} can be canceled by
a gauge transformation of $z_\alpha\rightarrow iz_\alpha$. Therefore
the state obtained by condensing $u_i v_x$ does not break the
translational symmetry.

Because of the gauge covariance of the bound-state operator, we need
to study its PSG property to fully understand the symmetries it
has. The symmetry transformations of the CP(1) field, the skyrmion,
and spinon pair operators are summarized in Table~\ref{tab:psg-field}.
A summary of symmetry transformations of the CP(1) field can be found
in Ref.~\onlinecite{rkk3}, and the symmetry transformations of
skyrmion operators are explained in Ref.~\onlinecite{senthil2004a}.

\begin{table}[htbp]
  \centering
  \begin{tabular*}{.7\textwidth}{@{\extracolsep{\fill} }cccccc}
    \hline\hline
    & $T_x$ & $T_y$ & $R_{\pi/2}$ & $I_x$ & $\mathcal{T}$\\
    \hline
    $z_\alpha$ & $\epsilon_{\alpha\beta}z_\beta^\ast$ &
     $\epsilon_{\alpha\beta}z_\beta^\ast$ & $z_\alpha$ &
     $z_\alpha$ &  $\epsilon_{\alpha\beta}z_\beta^\ast$\\
     $u_x$ & $u_x^\ast$ & $u_x^\ast$ & $u_y$ & $-u_x$ & $u_x^\ast$\\
     $u_y$ & $u_y^\ast$ & $u_y^\ast$ & $-u_x$ & $u_y$ & $u_y^\ast$\\
     $v_x$ & $-v_x$ & $v_x$ & $v_y$ & $-v_x$ & $v_x$\\
     $v_y$ & $v_y$ & $-v_y$ & $-v_x$ & $v_y$ & $v_y$\\
     $v$ & $-iv^\ast$ & $iv^\ast$ & $iv$ & $-v^\ast$ & $v^\ast$\\
     $f_x=u_xv_x$ & $-f_x^\ast$ & $f_x^\ast$ & $f_y$ & $f_x$ & $f_x^\ast$\\
     $g_x=u_xv_y$ & $-g_x^\ast$ & $g_x^\ast$ & $g_y$ & $-g_x$ & $g_x^\ast$\\
     \hline\hline
  \end{tabular*}
  \caption{Symmetry transformations of fields in the compact CP(1)
    model. Different columns represent actions of corresponding
    symmetry operations. $T_x$ and $T_y$: translations
    by one lattice spacing along $x$ and $y$ directions, respectively;
    $R_{\pi/2}$: 90-degree rotation about a lattice site; $I_x$:
    reflection about the axis of $x=0$; $\mathcal T$: time-reversal
    operation. $z_\alpha$ are the spinon fields in the CP(1) model in
    Eq.~\eqref{eq:cp1}, and its symmetry transformations are
    summarized in Ref.~\onlinecite{rkk3}; $u_{x,y}$ are the spinon
    pair operators defined in Eq.~\eqref{eq:umu}; $v$, $v_x$, and
    $v_y$ are skyrmion and VBS order parameters\cite{senthil2004a}
    defined in Eq.~\eqref{eq:vvxy}. $f_x$ and $g_x$ are two
    nematic bound-state operators defined in Eq.~\eqref{eq:fg}, and
    $f_y=u_yv_y$, $g_y=-u_yv_x$ are corresponding operators obtained
    after rotation.}
  \label{tab:psg-field}
\end{table}

Our aim is to find a bilinear form of $u$ and $v$ fields that is
invariant [up to a U(1) gauge transformation] under all symmetry
operations. However, this cannot be achieved, as $R_{\pi/2}$ and $T_x$
do not commute. In other words, condensing a bound state of
skyrmion/antiskyrmion and spinon pair will break either the
reflectional symmetry or the rotational symmetry. It is more natural
that we choose to break the rotational symmetry, as breaking the
translation enlarges the unit cell and allows the possibility of a
trivial paramagnetic ground state\cite{Hastings2004}. In the rest of
the paper we will consider only Z$_2$ spin liquid states where the
$C_4$ rotational symmetry of the square lattice is broken down to
$C_2$. In other words, the spin liquid states we obtain in this paper
are nematic spin liquid states. The possibility of obtaining a nematic
Z$_2$ spin liquid state in the $J_1$-$J_2$ model on a square lattice will be
discussed in more details in Sec~\ref{sec:conclusion}.

Finally, we fix the form of bound-state operator by considering the
requirement of reflection symmetry. The square lattice has reflection
symmetries with respect to both the $x$ and $y$ axes, and the diagonal
direction of $x\pm y$. When the four-fold rotation symmetry is broken,
only one set of reflection symmetries can be preserved. Here we
consider states with reflection symmetries about the $x$ and $y$ axes,
since these states have the same lattice symmetry as the $(0, \pi)$
N\'eel state at large $J_2/J_1$\cite{Jiang2012, Wang2011}. According
to Table~\ref{tab:psg-field}, the reflection symmetry changes $v$ to
its complex conjugate, so it turns a skyrmion into an
antiskyrmion. Therefore, to have a reflection symmetric condensate,
the order parameter needs to be a linear combination of spinon pair--
skyrmion bound state and spinon pair--antiskyrmion bound state. We
can show that there are two possibilities that satisfy all the
symmetries except rotation:
\begin{equation}
  \label{eq:fg}
  f_x = u_xv_x, g_x = u_yv_x.
\end{equation}
The symmetry transformations of these two fields are also summarized
in Table~\ref{tab:psg-field}. Under all symmetry transformations
except $R_{\pi/2}$, the two bound-state operators either are invariant
or become their complex conjugates, and they may also acquire a minus
sign. Using the U(1) gauge invariance, the phase of the bound-state
condensate can be fixed to be real, and the extra minus sign can also
be canceled by a U(1) gauge transformation. Therefore the states
obtained by condensing either $f_x$ or $g_x$ are nematic spin liquid
states that preserve all other symmetries listed in
Table~\ref{tab:psg-field}.

\section{Phase transition to Z$_2$ spin liquid state.}
\label{sec:eff}

In this section we discuss the phase transition to the Z$_2$ spin
liquid state and the low-energy excitations in the spin liquid
state. We will show that the Z$_2$ spin liquid state can be reached
from the deconfined quantum criticality by proliferating the spinon
pair--skyrmion/antiskyrmion bound states. Moreover, the vortex
excitations of the bound-state condensate become the vison
excitations in the Z$_2$ spin liquid state.

In the theory of the deconfined quantum criticality, killing the
N\'eel order in a spin-$\frac12$ system on square lattice brings the
system to the deconfined quantum critical point, which is described by
the noncompact CP(1) model. Away from the critical point, the
four-skyrmion tunneling events become a dangerously irrelevant
perturbation that drives the system into a VBS phase. This phase
transition can be described by the following effective Lagrangian:
\begin{equation}
  \label{eq:cp1-v4}
  \mathcal L = \frac1g\sum_\alpha|(\partial_\mu-ia_\mu)z_\alpha|^2
  +\lambda_v\left(v^4+v^{\dagger4}\right),
\end{equation}
where the $\lambda_v$ term represents four-skyrmion tunneling events.

Similarly, one can go from the deconfined quantum critical point to
the Z$_2$ spin liquid phase with the bound-state operator as another
dangerously irrelevant perturbation. Without losing generality, we
consider condensing $f_x$ as an example. The operator $f_x$ can be
decomposed into two fields describing bound states of spinon pair plus
skyrmion or antiskyrmion, respectively:
\begin{equation}
  \label{eq:f+-}
  f_x=\frac12(f_x^++f_x^-),\quad
  f_x^+=e^{-i\frac\pi4}v^\dagger u_x,\quad
  f_x^-=e^{i\frac\pi4}vu_x.
\end{equation}
As bound states, the gauge charge and flux carried by $f_x^\pm$ are the
sum of gauge charges carried by the spinon pair and the sum of gauge
flux carried by the skyrmion (or antiskyrmion). Hence $f_x^\pm$
carries gauge charge $2e$ and gauge flux $\pm2\pi$. In the CP(1)
model, the gauge charge is conserved, while the flux is conserved
modular $8\pi$, as skyrmion number is conserved modular
four. Therefore using the symmetry transformations listed in
Table~\ref{tab:psg-field} we see that the following Lagrangian with a
quartic term of bound-state operator is allowed by all lattice
symmetries and gauge charge and flux conservations,
\begin{equation}
  \label{eq:cp1-f4}
  \mathcal L = \frac1g\sum_\alpha|(\partial_\mu-ia_\mu)z_\alpha|^2
  +\lambda_f(f_x^{+2}f_x^{-\ast2}+\text{H.c.}).
\end{equation}
At the deconfined quantum critical point, the $f_x^\pm$ fields are
gapless as both spinon pair and skyrmion/antiskyrmion fields are
gapless. When we move away from the critical point, the $\lambda_f$
term in Eq.~\eqref{eq:cp1-f4} becomes relevant and leads to the
bound-state condensation. To be precise, this quartic term pins the
phases of $f_x^\pm$ fields, which breaks the U(1) gauge symmetry in
the CP(1) down to Z$_2$ and breaks the fourfold rotational
symmetry. We leave the study of the renormalization group flow of this
new quartic term to future works and only assume that such a
scenario of deconfined criticality is possible.  In the rest of this
section we discuss the low-energy excitations in the phase obtained
through bound-state condensation and argue that it is a gapped spin
liquid state with Z$_2$ topological order.

\begin{table}[htbp]
  \centering
  \begin{tabular*}{.7\textwidth}{@{\extracolsep{\fill} }ccc}
    \hline\hline
    Excitation & Gauge charge & Gauge flux \\
    \hline
    $z_\alpha$ & $e$ & 0 \\
    $v$ & 0 & $2\pi$ \\
    $f_x^\pm$ & $2e$ & $\mp2\pi$ \\
    Vortex of $f_x^\pm$ & $\mp e/2$ & $\pi/2$\\
    \hline\hline
  \end{tabular*}
  \caption{Gauge charge and gauge flux assignments of low-energy
    excitations. In the table $z_\alpha$ is spinon excitations in the
    CP(1) model, $v$ is skyrmion excitation, and $f_x\pm$ is the bound
    state of spinon pair and antiskyrmion/skyrmion defined in
    Eq.~\eqref{eq:f+-}.}
  \label{tab:gcgf}
\end{table}

As we are condensing the bound state of spinon pair and skyrmion, the
spinon excitations remain well defined in the condensed phase. Since
the condensate carries gauge flux $\pm2\pi$, the spinons are
gapped. Therefore in the condensed phase there are spin-$\frac12$
spinons carrying gauge charge $e$. On the other hand, in the condensed
phase there are also vortex excitations of the bound-state condensate.
Near the aforementioned critical point we have two condensates of
$f_x^\pm$, because the relative phase of the two is allowed to
fluctuate due to the irrelevance of the fourfold rotational lattice
anisotropy at the deconfined quantum critical point. Consequently,
there exist two types of topological excitations that are $2\pi$
vertices of the two condensates. The gauge charge and flux carried by
these excitations can be worked out by considering the mutual
statistics between the bound-state operators and their vortices: there
is a $2\pi$ Berry phase if we move an $f_x^\pm$ bound state
quasiparticle around the corresponding vortex, and there is no Berry
phase if we move an $f_x^\pm$ bound state around the vortex of the
opposite condensate $f_x^\mp$. Using this condition and the gauge
charge/flux assignment of $f_x^\pm$, we can derive the following gauge
charge/flux assignment of the vortices: the vortex of $f_x^+$ carries
gauge charge $-e/2$ and gauge flux $\pi/2$, and the vortex of $f_x^-$
carries carries gauge charge $e/2$ and gauge flux $\pi/2$. These
results are listed in Table~\ref{tab:gcgf}. Near the critical point
there are vortex excitations of $f_x^\pm$ carrying fractionalized
gauge charge and flux. However, when we move away from the critical
point into the bound state condensed phase, the phases of $f_x^\pm$
are locked by the quartic term in Eq.~\eqref{eq:cp1-f4} and there is
only one condensate of the linear combination of $f_x^\pm$ as shown in
Eq.~\eqref{eq:f+-}. Therefore the vortices of $f_x^\pm$ are confined
together and the bound state of two $f_x^\pm$ vortices carries no
gauge charge and gauge flux of $\pi$. In conclusion, in the bound
state condensed phase there are two types of low-energy excitations:
spinons carrying gauge charge $e$ and bound state of $f_x^\pm$
vortices carrying gauge flux $\pi$, and they see each other as $\pi$
flux. Therefore these two types of excitations can be treated as
spinon and vison excitations in a Z$_2$ spin liquid state, and
consequently the phase we get by condensing a spinon
pair--skyrmion/antiskyrmion bound state is a gapped spin liquid state
with Z$_2$ topological order.

Moreover, from this analysis one can see that both spinon and vison
gaps close at the critical point: The spinon gap closes since the
spinon condenses to form the N\'eel order as we go across the critical
point; the vison gap closes because the vortex core energy vanishes as
the stiffness of the $f_x^\pm$ condensates vanishes at the critical
point. This is consistent with the findings in the numerical
studies\cite{Jiang2012, Wang2011} that the gaps of spin-singlet and
spin-triplet excitations close as one approaches the quantum critical
point from the spin liquid side, and that both spin-spin and
dimer-dimer correlations have power-law behavior at the critical
point.

\section{Schwinger boson mean field state.}
\label{sec:sbmf}

In this section we construct a microscopic description of the nematic
spin liquid state obtained by condensing bound-state operator using
the Schwinger boson representation. The Schwinger boson method has
been used to study different spin models. Particularly, the nearest
neighbor Heisenberg model on square lattice has been studied using a
U(1) Schwinger boson spin liquid theory\cite{Auerbach1988,
  arovas1988}. Models with frustrations, like the $J_1$-$J_2$ model,
can be studied using a Z$_2$ Schwinger boson spin liquid
theory\cite{Flint2009}. In both cases, the Schwinger boson
representation introduces fractionalized spinons and emergent gauge
fields. Therefore, different projective ground state wave functions
have different topological orders which can be classified using their
PSG. Here we construct the particular mean field Hamiltonian that
gives the projective ground state corresponding to the spin liquid
which we obtain by the effective theory, by matching the PSG of the
mean field Hamiltonian to the PSG obtained in
Table~\ref{tab:psg-field}.

In the Schwinger boson representation, the spin degree of freedom is
expressed using two flavors of bosons carrying spin-$\frac12$,
\begin{equation}
  \label{eq:swb}
  \bm S_i = a_{i\alpha}^\dagger \bm\sigma_{\alpha\beta}a_{i\beta},
\end{equation}
where $\bm\sigma$ is a vector formed by the three Pauli matrices,
$\alpha,\beta$ are spin indices taking values of up and down, and
$a_{i\alpha}$ are Schwinger boson operators carrying
spin-$\frac12$. To relate the Schwinger boson representation to the
CP(1) model discussed in Sec.~\ref{sec:bs}, we adapt the notation in
Ref.~\onlinecite{rs1} where the Schwinger boson operator is redefined on
sublattice B as the following,
\begin{equation}
  \label{eq:swbb}
  b_{i\alpha} = \left\{\begin{matrix}
        a_{i\alpha},&i\in A,\\
        \epsilon_{\alpha\beta}a_{i\beta}^\dagger,&i\in B,
      \end{matrix}\right.
\end{equation}
where $\epsilon_{\alpha\beta}$ is the total antisymmetric
tensor. After this canonical transformation, the operator
$b_{i\alpha}$ is related to the physical spin operator as $(-1)^i\bm
S_i = b_{i\alpha}^\dagger \bm\sigma_{\alpha\beta}b_{i\beta}$, which
has a similar form as Eq.~\eqref{eq:zn}. Hence one can view the CP(1)
field $z_\alpha$ as the long-wavelength mode of $b_{i\alpha}$.

We start with a U(1) spin liquid state that corresponds to the
deconfined quantum critical point described by the CP(1) model. Such
state can be given by the following mean field Hamiltonian that
contains a uniform hopping term on nearest-neighbor
bonds\cite{Read1990},
\begin{equation}
  \label{eq:Hmf-u1}
  H_{\text{MF}}^{\text{nn}} = -P\sum_{\left<ij\right>}
  \left(b_{i\alpha}^\dagger b_{j\alpha} + \text{H.c.}\right),
\end{equation}
where $P$ is a mean field order parameter representing the hopping
matrix element on nearest-neighbor bonds. This mean field Hamiltonian
is invariant under U(1) gauge transformation $b_{i\alpha}\rightarrow
b_{i\alpha}e^{i\theta}$, and hence it is coupled to an emergent U(1) gauge
field. Moreover, the symmetry transformation of the spinon operator
$b_{i\alpha}$, as summarized in Table~\ref{tab:syswb}, is the same as
the CP(1) spinon field $z_\alpha$\cite{rkk3}. Consequently the U(1)
spin liquid state described here using Schwinger bosons represents the
same deconfined quantum critical point as in the case of the CP(1)
model in Eq.~\eqref{eq:cp1}, and the low-energy mode of $b_{i\alpha}$
corresponds to $z_\alpha$.

\begin{table}[htbp]
  \centering
  \begin{tabular*}{.7\textwidth}{@{\extracolsep{\fill} }cccccc}
    \hline\hline
    & $T_x$ & $T_y$ & $R_{\pi/2}$ & $I_x$ & $\mathcal{T}$\\
    \hline
    $b_{i\alpha}$ & $\epsilon_{\alpha\beta}b_{j\beta}^\ast$ &
     $\epsilon_{\alpha\beta}b_{j\beta}^\ast$ & $b_{j\alpha}$ &
     $b_{j\alpha}$ &  $\epsilon_{\alpha\beta}b_{j\beta}^\ast$\\
     \hline\hline
  \end{tabular*}
  \caption{Symmetry transformations of spinon in Schwinger boson mean
    field state\cite{rkk3}.}
  \label{tab:syswb}
\end{table}

Next, we study Z$_2$ spin liquid states adjacent to the deconfined
quantum critical point. Naturally, such states can be constructed on
top of this U(1) spin liquid state. Motivated by the $J_1$-$J_2$
model, we consider adding the following pairing term on the diagonal
bonds, which can lower the mean-field energy due to the $J_2$ coupling
in the Hamiltonian,
\begin{equation}
  \label{eq:Aij}
  H_{\text{MF}}^{\text{nnn}} = \sum_{\langle\langle ij\rangle\rangle}
  \left(Q_{ij}^\ast \epsilon_{\alpha\beta}b_{i\alpha}b_{j\beta}+
    Q_{ij} \epsilon_{\alpha\beta}b_{i\alpha}^\dagger
    b_{j\beta}^\dagger\right),
\end{equation}
where $Q_{ij}$ is the mean-field order parameter representing pairing
on next-nearest-neighbor (or diagonal) bonds, and it is proportional
to the mean-field expectation value of the spinon pair operator,
\begin{equation}
  \label{eq:Aijop}
  Q_{ij}\propto\langle \hat{A}_{ij}\rangle,\quad
  \hat A_{ij} = \epsilon_{\alpha\beta}b_{i\alpha}b_{j\beta}.
\end{equation}
Such a pairing term breaks the U(1) gauge symmetry and therefore changes
the gauge fluctuation to Z$_2$ through the Higgs mechanism.

In Sec.~\ref{sec:bs}, the Z$_2$ spin liquid state is obtained by
condensing the bound-state operator defined in Eq.~\eqref{eq:fg}. In
analogy, the Z$_2$ spin liquid state described here using Schwinger
boson framework is obtained by condensing pairs of Schwinger boson
operators. Consequently, in order to realize the same Z$_2$ spin
liquid state using Schwinger bosons, we need to find the particular
form of the spinon pair operator that corresponds to the bound-state
operator. At first glance, this task is not trivial because the
bound-state operator carries a skyrmion quantum number, which is a
topological defect of the spin state. In the theory of the deconfined
quantum criticality, the skyrmion operator is related to the order
parameter of the VBS state using the argument that the two operators
transform in the same way under all symmetry transformations, and
therefore have the same scaling behavior near the critical
point\cite{senthil2004a}. Similarly, we can find the form of the
bound-state operator in terms of Schwinger boson operators by
comparing how they transform under symmetry operations. In our case,
we need to find a Schwinger boson pair operator that has not only the
same symmetry, but also the same PSG as the bound-state operator, as
both operators carry gauge charge $2e$ and are thus
gauge covariant. Moreover, having the same PSG suggests that the two
states have the same topological order, which is required if they are
indeed the same state.

\begin{figure}[htbp]
  \centering
  \subfigure[\label{fig:pattern:fx}$f_x$]{\includegraphics{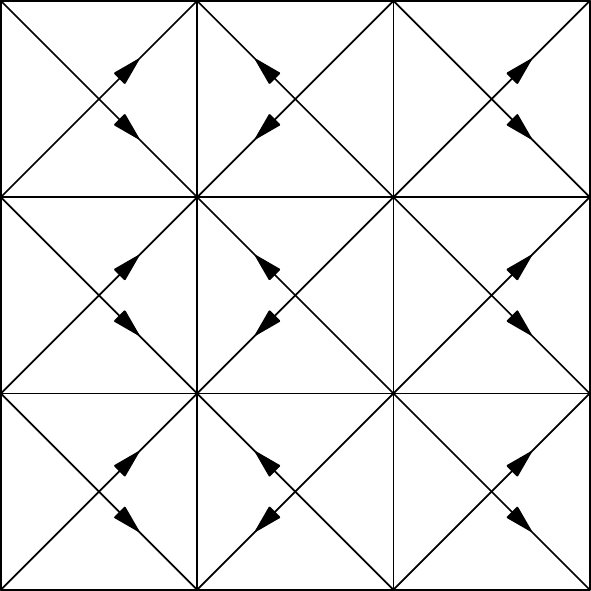}}
  \subfigure[\label{fig:pattern:gx}$g_x$]{\includegraphics{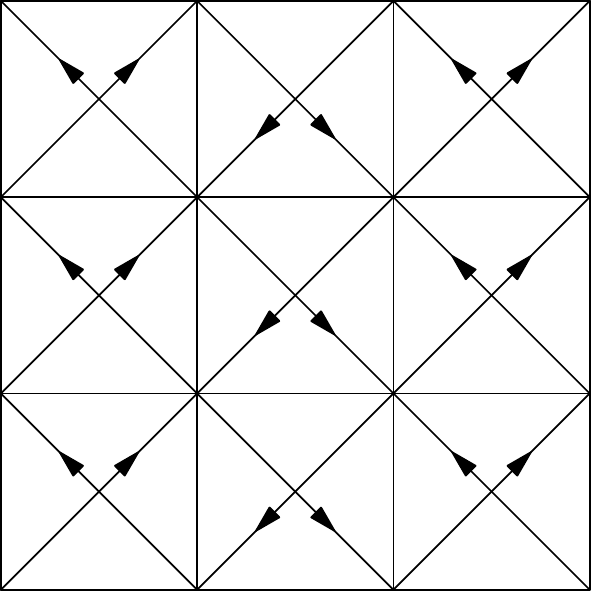}}
  \caption{Pattern of pairing order parameters $Q_{ij}$ in
    Eq.~\eqref{eq:Aij}. The arrows show the direction along which
    $Q_{ij}$ is positive. The two patterns correspond to spin liquid
    states obtained by condensing $f_x$ and $g_x$ as defined in
    Eq.~\eqref{eq:fg}, respectively.}
  \label{fig:pattern}
\end{figure}

The symmetry and topological order of the Z$_2$ spin liquid ground
state specified by the mean-field Hamiltonian in
Eqs.~\eqref{eq:Hmf-u1} and~\eqref{eq:Aij} are determined from
analyzing the PSG of the mean-field order parameters, particularly the
diagonal pairing order parameter $Q_{ij}$. Lattice symmetries and
time-reversal symmetry require that $Q_{ij}$ takes real values with
the same absolute value on all bonds, but it can have different signs
on different bonds. The sign of $Q_{ij}$ can be conveniently expressed
by specifying an orientation of the bond along which $Q_{ij}$ is
positive, as $Q_{ij}=-Q_{ji}$. Hence a pattern of $Q_{ij}$ can be
determined by specifying orientations of all diagonal bonds. Then the
PSG of this pattern can be worked out using the signs of $Q_{ij}$ and
the symmetry transformation of Schwinger boson operators listed in
Table~\ref{tab:syswb}. By matching the symmetry transformation with
the PSG of the bound-state operator listed in
Table~\ref{tab:psg-field}, we find the configuration of $Q_{ij}$ that
gives the same spin liquid state as obtained in Sec.~\ref{sec:bs}
by condensing $f_x$ and $g_x$ operators, and the configurations we
find are plotted in Fig.~\ref{fig:pattern}.

\section{Variational Monte Carlo study.}
\label{sec:vmc}

In this section we study the ground-state wave function of the
Schwinger boson projective ansatz using the variational Monte Carlo
(VMC) method. Here our primary goal is to illustrate that the
projective ansatz we propose based on the effective theory analysis
has a relatively low variational energy and is a possible candidate
state. Due to the sign problem in the VMC simulation, our study cannot
determine whether the Schwinger boson projective ansatz is the ground
state of the $J_1$-$J_2$ model.

Applying a Gutzwiller projection on mean-field ground-state wave
functions is a commonly used technique to improve the mean-field
results\cite{GROS1989}, and such a projection can be evaluated using the
VMC method. While being a popular technique to study fermionic
projective ansatzes, the VMC method is hard to apply to Schwinger
boson wave functions due to the difficulty of calculating
permanents\cite{Tay2011}.

Here we use an alternative VMC method that is based on the
non-orthogonal valence bond basis, which is first introduced by
\citet{Liang1988}. The Schwinger boson mean-field ground-state wave
function can be easily written in the valence bond basis. Following
the notation in Ref.~\onlinecite{Sandvik2010}, the wave function has
the following form:
\begin{equation}
  \label{eq:PsiVr}
  |\Psi\rangle = \sum_{V_r}w(V_r)|V_r\rangle,
\end{equation}
where $V_r$ denotes different spin-singlet valence bond covering
configurations,
\begin{equation}
  \label{eq:Vr}
  |V_r\rangle = |(a_1^r,b_1^r),(a_2^r,b_2^r),\ldots (a_{N/2}^r,b_{N/2}^r)\rangle,
\end{equation}
with $a_i^r$ and $b_i^r$ denoting the lattice sites of the $i$th
valence bond, and we assume that the weight of each configuration is
given by a product of the weight of each bond,
\begin{equation}
  \label{eq:WVr}
  w(V_r)=\prod_iw(a_i^r, b_i^r).
\end{equation}

Using the $a_{i\alpha}$ Schwinger boson operators, the mean field
Hamiltonian in Eq.~\eqref{eq:Hmf-u1} and~\eqref{eq:Aij} has the
following form,
\begin{equation}
  \label{eq:hmf-a}
  H_{\text{MF}} = -\sum_{\langle ij\rangle}P_{ij}\left(
    a_{i\alpha}^\dagger a_{j\alpha}+\text{H.c.}\right)
  +\sum_{\langle\langle ij\rangle\rangle}\left(
    Q_{ij}^\ast\epsilon_{\alpha\beta}a_{i\alpha\beta}
    +\text{H.c.}\right),
\end{equation}
and contains pairing terms on both nearest-neighbor and diagonal
bonds.  As a result, after applying the Gutzwiller projection, the
Schwinger boson mean-field wave function can be written in forms of
Eq.~\eqref{eq:PsiVr} with weights $w(V_r)$ determined from the
mean-field Hamiltonian\cite{Tay2011}. However, here we use a more
general form of variational wave function where we assume that the
absolute value of the weights depends only on the Manhattan distance
of the bond and use weights of different bonds instead of the
parameters in the mean-field Hamiltonian as variational parameters.

On the other hand, the sign of the weights is determined from the
projective symmetry group of the mean-field ansatz. For a U(1) spin
liquid ansatz, the ground state in Eq.~\eqref{eq:hmf-a} contains only
valence bond pairings between two sublattices and the weights are all
positive(the orientation of bonds is chosen to be pointing from
sublattice A to sublattice B\cite{Liang1988}). Therefore the VMC does
not have any sign problem and converges rapidly. However, the Z$_2$
spin liquid state obtained after condensing the spinon pair operator
in Eq.~\eqref{eq:Aijop} does create the sign problem in the VMC
calculation. However, for a finite size the sign problem can be
overcome by brutal force if the diagonal pairing amplitude is small
enough.

We perform the VMC calculation using the improved loop update
algorithm\cite{Sandvik2010}. To study the U(1) spin liquid state, we
go beyond a simple mean-field ansatz of Eq.~\eqref{eq:hmf-a} and allow
pairings on all inter-sublattice bonds. We assume that the weights of
bonds depends only on the Manhattan length of the bonds and use the
weights as variational parameters. On a 32-by-32 sites system we
obtain a ground state energy of $-0.4893(2)J_1$ per site with
$J_2=0.5J_1$, and $-0.4748(2)J_1$ with $J_2=0.55J_1$. Comparing to the
ground state energy of $-0.4943J_1$ for $J_2=0.5J_1$ and $-0.4844J_1$
for $J_2=0.55J_1$ obtained in Ref.~\onlinecite{Wang2011}, this
suggests that a bosonic U(1) spin liquid state is a reasonable
starting point in understanding the spin liquid phase in the
$J_1$-$J_2$ model. The bond weights $w(a, b)$ obtained from the
variational calculation decay exponentially as the length of the bond
increases, indicating that the spin liquid state has short-range
spin-spin correlation\cite{Liang1988}. Here we emphasize that this
wave function corresponds to the parent critical U(1) state described
by the critical CP(1) model or the U(1) Schwinger boson ansatz, not
the gapped Z$_2$ spin liquid state, which we will discuss briefly
later (hence we do not expect this wave function to give a low
variational energy as compared to other numerical
methods). Particularly, this wave function contains only short-ranged
intersublattice bonds and therefore has a U(1) topological order. As
a result, it has a critical dimer-dimer correlation\cite{Tang2011}.

Starting from this critical U(1) spin liquid state, we obtain a Z$_2$
spin liquid state by adding a small weight of diagonal pairing, and the
signs of the diagonal pairing are given by the ansatz shown in
Fig.~\ref{fig:pattern}. The numerical results are listed in
Table~\ref{tab:j1xy}. For either ansatz, we observe that there is no
change in the ground-state energy within our statistical errors, but
for the $f_x$ ansatz, introducing the diagonal pairing creates
anisotropy in nearest-neighbor spin-spin correlation. In other words,
the Z$_2$ spin liquid state with a diagonal pairing does not improve
the energy. Our numerical study suggests that the bosonic nematic spin
liquid state has a low ground-state energy as a variational state, but
whether it is the ground state of the $J_1$-$J_2$ model cannot be
concluded from our variational calculation. On the other hand, the
anisotropic $\bm S_i\cdot \bm S_j$ on nearest-neighbor bonds implies
that this nematic spin liquid state has a lower energy in an
anisotropic $J_{1x}$-$J_{1y}$-$J_2$ model, where the nearest-neighbor
antiferromagnetic interactions in the $x$ and $y$ directions are
different: $J_{1x}\neq J_{1y}$. There have been numerical studies on
this $J_{1x}$-$J_{1y}$-$J_2$ model\cite{Bishop2008} that show the
existence of an intermediate nonmagnetic phase between the N\'eel
state and another antiferromagnetic phase with a $(\pi, 0)$ order for
a finite range of $J_{1x}/J_{1y}$ around 1. This suggests that such
a spin liquid phase also exists when $J_{1x}\neq J_{1y}$, and the
nematic Schwinger boson projective wave function we study in this work
may describe such a spin liquid state in the anisotropy
$J_{1x}$-$J_{1y}$-$J_2$ model.


\begin{table}[htbp]
  \centering
  \begin{tabular*}{.7\textwidth}{@{\extracolsep{\fill} }lcc}
    \hline\hline
    Wave function & Energy per site/$J_1$ & $|(C_x-C_y)/(C_x+C_y)|$ \\
    \hline
    $w_d=0$ & $-0.489281(1)$ & 0 \\
    $f_x$, $w_d=0.005$ & $-0.489280(1)$ & 0.000045(10) \\
    $f_x$, $w_d=0.01$ & $-0.489284(3)$ & 0.000184(26) \\
    $g_x$, $w_d=0.005$ & $-0.489282(1)$ & 0.000017(10) \\
    $g_x$, $w_d=0.01$ & $-0.489281(3)$ & 0.000023(26) \\
    \hline\hline
  \end{tabular*}
  \caption{Energy and anisotropy of nearest neighbor spin-spin
    correlation of variational wave functions. In the first column,
    $w_d$ denotes the weight of the diagonal bonds defined in
    Eq.~\eqref{eq:WVr}, relative to the weight of nearest-neighbor
    bonds. $f_x$ and $g_x$, respectively, denote the pattern shown in the two
    subfigures in Fig.~\ref{fig:pattern}. The second
    column shows the energy per site in units of $J_1$, and the third
    column shows the anisotropy of nearest-neighbor spin-spin
    correlations, where $C_{x,y}=\langle\bm S_i\cdot\bm
    S_{i+x,y}\rangle$ is the nearest-neighbor spin-spin correlation in
    $x$ and $y$ directions, respectively. The number in the
    parenthesis shows the standard error. Note that the energies
    listed here have smaller errors compared to the ground-state
    energy $-0.4893(2)$ given in the main text, because the errors
    listed here contain only the statistical errors in the Monte
    Carlo simulations, whereas the main error in the ground-state
    energy data provided in the main content comes from minimizing the
    energy of trial wave function.}
  \label{tab:j1xy}
\end{table}

\section{Conclusions}
\label{sec:conclusion}

In this paper we have discussed a possible scenario of obtaining a
Z$_2$ spin liquid phase from the N\'eel phase in a spin-$\frac12$
system on a square lattice through a continuous phase transition by
condensing a bound state of spinon pair and skyrmion excitations. The
symmetry of the spin liquid state is studied using PSG analysis. While
condensing the skyrmion itself breaks the translational symmetry, the
bound-state condensation does not break this symmetry and leads to a
translational symmetric spin liquid state. Near the critical point,
the vortices of the condensate carry fractionalized gauge charge and
flux, but they are confined in the spin liquid phase and are combined
to form vison excitations in the Z$_2$ gauge theory. Moreover, we can
describe the Z$_2$ spin liquid state using a Schwinger boson
projective wave function and the bound-state operator maps to a
pairing operator on diagonal bonds with a certain PSG. We calculate
the ground-state energy of the Schwinger boson projective wave
function using the variational Monte Carlo method and find that it has
a relatively low energy. The spin liquid state we obtain has the Z$_2$
topological order, and therefore the entanglement entropy contains the
universal constant $\gamma=\ln 2$, which is consistent with the
observations in numerical studies\cite{Jiang2012, Wang2011}.

The spin liquid state we obtain in this work is nematic as it has all
translational symmetries of the square lattice but breaks the
fourfold rotational symmetry down to twofold. The result that we
could not find a rotational symmetric spin liquid state is consistent
with previous studies on slave-particle constructions of spin liquid
states on the square lattice. On one hand, using the Schwinger boson
framework, nematic spin liquid states have been proposed on a square
lattice\cite{rs1,Sachdev1991}, and have been used to study the
$J_1$-$J_2$ model\cite{Flint2009}. Moreover, the PSG
analysis\cite{Yang2012} shows that all bosonic spin liquid states that
have zero-flux hopping on nearest-neighbor bonds and nonvanishing
pairing on diagonal bonds are nematic. In other words, all Z$_2$ spin
liquid states obtained by adding pairing on diagonal bonds on top of
the U(1) spin liquid state are nematic. On the other hand, the PSG
analysis on fermionic spin liquid states\cite{wenpsg} shows that there
is no rotational symmetric gapped Z$_2$ spin liquid state adjacent to
the $\pi$-flux U(1) spin liquid state. In summary, neither a bosonic nor
fermionic slave particle framework can describe a rotational symmetric
gapped Z$_2$ spin liquid state that can be connected to the N\'eel
state through a continuous phase transition. Furthermore, we note that
a similar lattice symmetry-breaking spin liquid state is proposed for
the kagome lattice Heisenberg model\cite{Capponi2013}. However, on the
square lattice the lattice symmetry breaking plays a more crucial role
in the Z$_2$ spin liquid state, because without such symmetry breaking
the spin liquid state would be coupled to a U(1) gauge field instead,
which would make it unstable in two dimensions\cite{Read1990}.

One key result of this theoretical work is that on the square lattice,
the gapped spin liquid state obtained through a direct second-order
phase transition from the N\'eel state is a nematic spin liquid state
that breaks the four-fold rotational symmetry. Such symmetry breaking
is neither observed nor ruled out in numerical studies of the
$J_1$-$J_2$ model. On one hand, the system studied in
Ref.~\onlinecite{Jiang2012} using the density matrix renormalization
group (DMRG) method is a ladder system and does not have the
rotational symmetry to begin with. On the other hand, in the work of
\citet{Wang2011} rotational symmetry of the ground state was not
explicitly checked. We hope the rotational symmetry of the spin liquid
state can be clarified by future numerical studies. Moreover, recent
numerical studies using DMRG\cite{Gong2013a} and VMC
methods\cite{Hu2013} provide evidence for a gapless spin liquid
state. Therefore we hope future numerical studies can resolve this
controversy and determine whether our critical theory can be applied
to the $J_1$-$J_2$ model on a square lattice.

Even though the nematic spin liquid state we have
proposed may not describe the ground state of the $J_1$-$J_2$ model,
it still might be realized in a model that lacks $C_4$ lattice rotational
symmetry, as suggested by our variational study described in
Sec.~\ref{sec:vmc}. We note that our theoretical analysis in
Sec.~\ref{sec:bs}--\ref{sec:sbmf} also applies to an anisotropic
model. Particularly, the symmetry transformations listed in
Table~\ref{tab:psg-field} generate all lattice symmetry operations of
an anisotropic square lattice if one replaces the rotation $R_{\pi/2}$
by $R_\pi=R_{\pi/2}^2$. Hence the same novel quantum critical point
between the N\'eel and the Z$_2$ spin liquid state also exists in an
anisotropic model. Therefore it will be interesting to study the
anisotropic $J_{1x}$-$J_{1y}$-$J_2$ model to see if the anisotropy
helps to stabilize the nematic spin liquid state found in this
work.  

\begin{acknowledgments}
  We thank L. Wang, C. Xu,  T. Senthil, E.-G. Moon, and S. Sachdev for useful
  discussions.
  YQ is supported by the National Natural Science Foundation of China
  through grant No. 11104154 and by the National Basic Research
  Program of China through Grant No. 2011CBA00108.
  ZCG is supported in part by the NSF Frontiers Center with support from the Gordon and Betty Moore Foundation.
\end{acknowledgments}


\begin{thebibliography}{37}%
\makeatletter
\providecommand \@ifxundefined [1]{%
 \@ifx{#1\undefined}
}%
\providecommand \@ifnum [1]{%
 \ifnum #1\expandafter \@firstoftwo
 \else \expandafter \@secondoftwo
 \fi
}%
\providecommand \@ifx [1]{%
 \ifx #1\expandafter \@firstoftwo
 \else \expandafter \@secondoftwo
 \fi
}%
\providecommand \natexlab [1]{#1}%
\providecommand \enquote  [1]{``#1''}%
\providecommand \bibnamefont  [1]{#1}%
\providecommand \bibfnamefont [1]{#1}%
\providecommand \citenamefont [1]{#1}%
\providecommand \href@noop [0]{\@secondoftwo}%
\providecommand \href [0]{\begingroup \@sanitize@url \@href}%
\providecommand \@href[1]{\@@startlink{#1}\@@href}%
\providecommand \@@href[1]{\endgroup#1\@@endlink}%
\providecommand \@sanitize@url [0]{\catcode `\\12\catcode `\$12\catcode
  `\&12\catcode `\#12\catcode `\^12\catcode `\_12\catcode `\%12\relax}%
\providecommand \@@startlink[1]{}%
\providecommand \@@endlink[0]{}%
\providecommand \url  [0]{\begingroup\@sanitize@url \@url }%
\providecommand \@url [1]{\endgroup\@href {#1}{\urlprefix }}%
\providecommand \urlprefix  [0]{URL }%
\providecommand \Eprint [0]{\href }%
\providecommand \doibase [0]{http://dx.doi.org/}%
\providecommand \selectlanguage [0]{\@gobble}%
\providecommand \bibinfo  [0]{\@secondoftwo}%
\providecommand \bibfield  [0]{\@secondoftwo}%
\providecommand \translation [1]{[#1]}%
\providecommand \BibitemOpen [0]{}%
\providecommand \bibitemStop [0]{}%
\providecommand \bibitemNoStop [0]{.\EOS\space}%
\providecommand \EOS [0]{\spacefactor3000\relax}%
\providecommand \BibitemShut  [1]{\csname bibitem#1\endcsname}%
\let\auto@bib@innerbib\@empty
\bibitem [{\citenamefont {Lee}\ \emph {et~al.}(2006)\citenamefont {Lee},
  \citenamefont {Nagaosa},\ and\ \citenamefont {Wen}}]{LNW2006Review}%
  \BibitemOpen
  \bibfield  {author} {\bibinfo {author} {\bibfnamefont {P.}~\bibnamefont
  {Lee}}, \bibinfo {author} {\bibfnamefont {N.}~\bibnamefont {Nagaosa}}, \ and\
  \bibinfo {author} {\bibfnamefont {X.-G.}\ \bibnamefont {Wen}},\ }\href
  {\doibase 10.1103/RevModPhys.78.17} {\bibfield  {journal} {\bibinfo
  {journal} {Rev. Mod. Phys.}\ }\textbf {\bibinfo {volume} {78}},\ \bibinfo
  {pages} {17} (\bibinfo {year} {2006})}\BibitemShut {NoStop}%
\bibitem [{\citenamefont {Zhang}\ and\ \citenamefont {Rice}(1988)}]{Zhang1988}%
  \BibitemOpen
  \bibfield  {author} {\bibinfo {author} {\bibfnamefont {F.~C.}\ \bibnamefont
  {Zhang}}\ and\ \bibinfo {author} {\bibfnamefont {T.~M.}\ \bibnamefont
  {Rice}},\ }\href {\doibase 10.1103/PhysRevB.37.3759} {\bibfield  {journal}
  {\bibinfo  {journal} {Phys. Rev. B}\ }\textbf {\bibinfo {volume} {37}},\
  \bibinfo {pages} {3759} (\bibinfo {year} {1988})}\BibitemShut {NoStop}%
\bibitem [{\citenamefont {Anderson}(1973)}]{Anderson1973}%
  \BibitemOpen
  \bibfield  {author} {\bibinfo {author} {\bibfnamefont {P.~W.}\ \bibnamefont
  {Anderson}},\ }\href {\doibase 10.1016/0025-5408(73)90167-0} {\bibfield
  {journal} {\bibinfo  {journal} {Mater. Res. Bull.}\ }\textbf {\bibinfo
  {volume} {8}},\ \bibinfo {pages} {153} (\bibinfo {year} {1973})}\BibitemShut
  {NoStop}%
\bibitem [{\citenamefont {Anderson}(1987)}]{Anderson1987}%
  \BibitemOpen
  \bibfield  {author} {\bibinfo {author} {\bibfnamefont {P.~W.}\ \bibnamefont
  {Anderson}},\ }\href {\doibase 10.1126/science.235.4793.1196} {\bibfield
  {journal} {\bibinfo  {journal} {Science}\ }\textbf {\bibinfo {volume}
  {235}},\ \bibinfo {pages} {1196} (\bibinfo {year} {1987})}\BibitemShut
  {NoStop}%
\bibitem [{\citenamefont {Dai}\ \emph {et~al.}(2012)\citenamefont {Dai},
  \citenamefont {Hu},\ and\ \citenamefont {Dagotto}}]{Dai2012}%
  \BibitemOpen
  \bibfield  {author} {\bibinfo {author} {\bibfnamefont {P.}~\bibnamefont
  {Dai}}, \bibinfo {author} {\bibfnamefont {J.}~\bibnamefont {Hu}}, \ and\
  \bibinfo {author} {\bibfnamefont {E.}~\bibnamefont {Dagotto}},\ }\href
  {\doibase 10.1038/nphys2438} {\bibfield  {journal} {\bibinfo  {journal} {Nat.
  Phys.}\ }\textbf {\bibinfo {volume} {8}},\ \bibinfo {pages} {709} (\bibinfo
  {year} {2012})}\BibitemShut {NoStop}%
\bibitem [{\citenamefont {Melzi}\ \emph {et~al.}(2001)\citenamefont {Melzi},
  \citenamefont {Aldrovandi}, \citenamefont {Tedoldi}, \citenamefont
  {Carretta}, \citenamefont {Millet},\ and\ \citenamefont {Mila}}]{Melzi2001}%
  \BibitemOpen
  \bibfield  {author} {\bibinfo {author} {\bibfnamefont {R.}~\bibnamefont
  {Melzi}}, \bibinfo {author} {\bibfnamefont {S.}~\bibnamefont {Aldrovandi}},
  \bibinfo {author} {\bibfnamefont {F.}~\bibnamefont {Tedoldi}}, \bibinfo
  {author} {\bibfnamefont {P.}~\bibnamefont {Carretta}}, \bibinfo {author}
  {\bibfnamefont {P.}~\bibnamefont {Millet}}, \ and\ \bibinfo {author}
  {\bibfnamefont {F.}~\bibnamefont {Mila}},\ }\href {\doibase
  10.1103/PhysRevB.64.024409} {\bibfield  {journal} {\bibinfo  {journal} {Phys.
  Rev. B}\ }\textbf {\bibinfo {volume} {64}},\ \bibinfo {pages} {024409}
  (\bibinfo {year} {2001})}\BibitemShut {NoStop}%
\bibitem [{\citenamefont {Read}\ and\ \citenamefont {Sachdev}(1991)}]{rs1}%
  \BibitemOpen
  \bibfield  {author} {\bibinfo {author} {\bibfnamefont {N.}~\bibnamefont
  {Read}}\ and\ \bibinfo {author} {\bibfnamefont {S.}~\bibnamefont {Sachdev}},\
  }\href {\doibase 10.1103/PhysRevLett.66.1773} {\bibfield  {journal} {\bibinfo
   {journal} {Phys. Rev. Lett.}\ }\textbf {\bibinfo {volume} {66}},\ \bibinfo
  {pages} {1773} (\bibinfo {year} {1991})}\BibitemShut {NoStop}%
\bibitem [{\citenamefont {Sachdev}\ and\ \citenamefont
  {Read}(1991)}]{Sachdev1991}%
  \BibitemOpen
  \bibfield  {author} {\bibinfo {author} {\bibfnamefont {S.}~\bibnamefont
  {Sachdev}}\ and\ \bibinfo {author} {\bibfnamefont {N.}~\bibnamefont {Read}},\
  }\href {\doibase 10.1142/S0217979291000158} {\bibfield  {journal} {\bibinfo
  {journal} {Int. J. Mod. Phys. B}\ }\textbf {\bibinfo {volume} {5}},\ \bibinfo
  {pages} {219} (\bibinfo {year} {1991})},\ \Eprint
  {http://arxiv.org/abs/cond-mat/0402109} {arXiv:cond-mat/0402109} \BibitemShut
  {NoStop}%
\bibitem [{\citenamefont {Flint}\ and\ \citenamefont
  {Coleman}(2009)}]{Flint2009}%
  \BibitemOpen
  \bibfield  {author} {\bibinfo {author} {\bibfnamefont {R.}~\bibnamefont
  {Flint}}\ and\ \bibinfo {author} {\bibfnamefont {P.}~\bibnamefont
  {Coleman}},\ }\href {\doibase 10.1103/PhysRevB.79.014424} {\bibfield
  {journal} {\bibinfo  {journal} {Phys. Rev. B}\ }\textbf {\bibinfo {volume}
  {79}},\ \bibinfo {pages} {014424} (\bibinfo {year} {2009})}\BibitemShut
  {NoStop}%
\bibitem [{\citenamefont {Jiang}\ \emph {et~al.}(2012)\citenamefont {Jiang},
  \citenamefont {Yao},\ and\ \citenamefont {Balents}}]{Jiang2012}%
  \BibitemOpen
  \bibfield  {author} {\bibinfo {author} {\bibfnamefont {H.-C.}\ \bibnamefont
  {Jiang}}, \bibinfo {author} {\bibfnamefont {H.}~\bibnamefont {Yao}}, \ and\
  \bibinfo {author} {\bibfnamefont {L.}~\bibnamefont {Balents}},\ }\href
  {\doibase 10.1103/PhysRevB.86.024424} {\bibfield  {journal} {\bibinfo
  {journal} {Phys. Rev. B}\ }\textbf {\bibinfo {volume} {86}},\ \bibinfo
  {pages} {024424} (\bibinfo {year} {2012})}\BibitemShut {NoStop}%
\bibitem [{\citenamefont {Wang}\ \emph {et~al.}()\citenamefont {Wang},
  \citenamefont {Gu}, \citenamefont {Verstraete},\ and\ \citenamefont
  {Wen}}]{Wang2011}%
  \BibitemOpen
  \bibfield  {author} {\bibinfo {author} {\bibfnamefont {L.}~\bibnamefont
  {Wang}}, \bibinfo {author} {\bibfnamefont {Z.-C.}\ \bibnamefont {Gu}},
  \bibinfo {author} {\bibfnamefont {F.}~\bibnamefont {Verstraete}}, \ and\
  \bibinfo {author} {\bibfnamefont {X.-G.}\ \bibnamefont {Wen}},\ }\href
  {http://arxiv.org/abs/1112.3331} {\ }\Eprint {http://arxiv.org/abs/1112.3331}
  {arXiv:1112.3331} \BibitemShut {NoStop}%
\bibitem [{\citenamefont {Senthil}\ \emph
  {et~al.}(2004{\natexlab{a}})\citenamefont {Senthil}, \citenamefont
  {Vishwanath}, \citenamefont {Balents}, \citenamefont {Sachdev},\ and\
  \citenamefont {Fisher}}]{senthil2004}%
  \BibitemOpen
  \bibfield  {author} {\bibinfo {author} {\bibfnamefont {T.}~\bibnamefont
  {Senthil}}, \bibinfo {author} {\bibfnamefont {A.}~\bibnamefont {Vishwanath}},
  \bibinfo {author} {\bibfnamefont {L.}~\bibnamefont {Balents}}, \bibinfo
  {author} {\bibfnamefont {S.}~\bibnamefont {Sachdev}}, \ and\ \bibinfo
  {author} {\bibfnamefont {M.~P.~A.}\ \bibnamefont {Fisher}},\ }\href {\doibase
  10.1126/science.1091806} {\bibfield  {journal} {\bibinfo  {journal}
  {Science}\ }\textbf {\bibinfo {volume} {303}},\ \bibinfo {pages} {1490}
  (\bibinfo {year} {2004}{\natexlab{a}})}\BibitemShut {NoStop}%
\bibitem [{\citenamefont {Senthil}\ \emph
  {et~al.}(2004{\natexlab{b}})\citenamefont {Senthil}, \citenamefont {Balents},
  \citenamefont {Sachdev}, \citenamefont {Vishwanath},\ and\ \citenamefont
  {Fisher}}]{senthil2004a}%
  \BibitemOpen
  \bibfield  {author} {\bibinfo {author} {\bibfnamefont {T.}~\bibnamefont
  {Senthil}}, \bibinfo {author} {\bibfnamefont {L.}~\bibnamefont {Balents}},
  \bibinfo {author} {\bibfnamefont {S.}~\bibnamefont {Sachdev}}, \bibinfo
  {author} {\bibfnamefont {A.}~\bibnamefont {Vishwanath}}, \ and\ \bibinfo
  {author} {\bibfnamefont {M.~P.~A.}\ \bibnamefont {Fisher}},\ }\href {\doibase
  10.1103/PhysRevB.70.144407} {\bibfield  {journal} {\bibinfo  {journal} {Phys.
  Rev. B}\ }\textbf {\bibinfo {volume} {70}},\ \bibinfo {pages} {144407}
  (\bibinfo {year} {2004}{\natexlab{b}})}\BibitemShut {NoStop}%
\bibitem [{\citenamefont {Sachdev}(2008)}]{Sachdev2008}%
  \BibitemOpen
  \bibfield  {author} {\bibinfo {author} {\bibfnamefont {S.}~\bibnamefont
  {Sachdev}},\ }\href {\doibase 10.1038/nphys894} {\bibfield  {journal}
  {\bibinfo  {journal} {Nat. Phys.}\ }\textbf {\bibinfo {volume} {4}},\
  \bibinfo {pages} {173} (\bibinfo {year} {2008})}\BibitemShut {NoStop}%
\bibitem [{\citenamefont {Xu}\ and\ \citenamefont {Sachdev}(2009)}]{xucoming}%
  \BibitemOpen
  \bibfield  {author} {\bibinfo {author} {\bibfnamefont {C.}~\bibnamefont
  {Xu}}\ and\ \bibinfo {author} {\bibfnamefont {S.}~\bibnamefont {Sachdev}},\
  }\href {\doibase 10.1103/PhysRevB.79.064405} {\bibfield  {journal} {\bibinfo
  {journal} {Phys. Rev. B}\ }\textbf {\bibinfo {volume} {79}},\ \bibinfo
  {pages} {064405} (\bibinfo {year} {2009})}\BibitemShut {NoStop}%
\bibitem [{\citenamefont {Moon}\ and\ \citenamefont {Xu}(2012)}]{Moon2012}%
  \BibitemOpen
  \bibfield  {author} {\bibinfo {author} {\bibfnamefont {E.-G.}\ \bibnamefont
  {Moon}}\ and\ \bibinfo {author} {\bibfnamefont {C.}~\bibnamefont {Xu}},\
  }\href {\doibase 10.1103/PhysRevB.86.214414} {\bibfield  {journal} {\bibinfo
  {journal} {Phys. Rev. B}\ }\textbf {\bibinfo {volume} {86}},\ \bibinfo
  {pages} {214414} (\bibinfo {year} {2012})}\BibitemShut {NoStop}%
\bibitem [{\citenamefont {Sandvik}(2010)}]{Sandvik2010a}%
  \BibitemOpen
  \bibfield  {author} {\bibinfo {author} {\bibfnamefont {A.~W.}\ \bibnamefont
  {Sandvik}},\ }\href {\doibase 10.1103/PhysRevLett.104.177201} {\bibfield
  {journal} {\bibinfo  {journal} {Phys. Rev. Lett.}\ }\textbf {\bibinfo
  {volume} {104}},\ \bibinfo {pages} {177201} (\bibinfo {year}
  {2010})}\BibitemShut {NoStop}%
\bibitem [{\citenamefont {Mambrini}\ \emph {et~al.}(2006)\citenamefont
  {Mambrini}, \citenamefont {L\"{a}uchli}, \citenamefont {Poilblanc},\ and\
  \citenamefont {Mila}}]{Mambrini2006}%
  \BibitemOpen
  \bibfield  {author} {\bibinfo {author} {\bibfnamefont {M.}~\bibnamefont
  {Mambrini}}, \bibinfo {author} {\bibfnamefont {A.}~\bibnamefont
  {L\"{a}uchli}}, \bibinfo {author} {\bibfnamefont {D.}~\bibnamefont
  {Poilblanc}}, \ and\ \bibinfo {author} {\bibfnamefont {F.}~\bibnamefont
  {Mila}},\ }\href {\doibase 10.1103/PhysRevB.74.144422} {\bibfield  {journal}
  {\bibinfo  {journal} {Phys. Rev. B}\ }\textbf {\bibinfo {volume} {74}},\
  \bibinfo {pages} {144422} (\bibinfo {year} {2006})}\BibitemShut {NoStop}%
\bibitem [{\citenamefont {Auerbach}\ and\ \citenamefont
  {Arovas}(1988)}]{Auerbach1988}%
  \BibitemOpen
  \bibfield  {author} {\bibinfo {author} {\bibfnamefont {A.}~\bibnamefont
  {Auerbach}}\ and\ \bibinfo {author} {\bibfnamefont {D.~P.}\ \bibnamefont
  {Arovas}},\ }\href {\doibase 10.1103/PhysRevLett.61.617} {\bibfield
  {journal} {\bibinfo  {journal} {Phys. Rev. Lett.}\ }\textbf {\bibinfo
  {volume} {61}},\ \bibinfo {pages} {617} (\bibinfo {year} {1988})}\BibitemShut
  {NoStop}%
\bibitem [{\citenamefont {Arovas}\ and\ \citenamefont
  {Auerbach}(1988)}]{arovas1988}%
  \BibitemOpen
  \bibfield  {author} {\bibinfo {author} {\bibfnamefont {D.~P.}\ \bibnamefont
  {Arovas}}\ and\ \bibinfo {author} {\bibfnamefont {A.}~\bibnamefont
  {Auerbach}},\ }\href {\doibase 10.1103/PhysRevB.38.316} {\bibfield  {journal}
  {\bibinfo  {journal} {Phys. Rev. B}\ }\textbf {\bibinfo {volume} {38}},\
  \bibinfo {pages} {316} (\bibinfo {year} {1988})}\BibitemShut {NoStop}%
\bibitem [{\citenamefont {Li}\ \emph {et~al.}(2012)\citenamefont {Li},
  \citenamefont {Becca}, \citenamefont {Hu},\ and\ \citenamefont
  {Sorella}}]{Li2012}%
  \BibitemOpen
  \bibfield  {author} {\bibinfo {author} {\bibfnamefont {T.}~\bibnamefont
  {Li}}, \bibinfo {author} {\bibfnamefont {F.}~\bibnamefont {Becca}}, \bibinfo
  {author} {\bibfnamefont {W.}~\bibnamefont {Hu}}, \ and\ \bibinfo {author}
  {\bibfnamefont {S.}~\bibnamefont {Sorella}},\ }\href
  {http://arxiv.org/abs/1205.3838
  http://link.aps.org/doi/10.1103/PhysRevB.86.075111} {\bibfield  {journal}
  {\bibinfo  {journal} {Phys. Rev. B}\ }\textbf {\bibinfo {volume} {86}},\
  \bibinfo {pages} {075111} (\bibinfo {year} {2012})}\BibitemShut {NoStop}%
\bibitem [{\citenamefont {Wen}(2002)}]{wenpsg}%
  \BibitemOpen
  \bibfield  {author} {\bibinfo {author} {\bibfnamefont {X.-G.}\ \bibnamefont
  {Wen}},\ }\href {\doibase 10.1103/PhysRevB.65.165113} {\bibfield  {journal}
  {\bibinfo  {journal} {Phys. Rev. B}\ }\textbf {\bibinfo {volume} {65}},\
  \bibinfo {pages} {165113} (\bibinfo {year} {2002})}\BibitemShut {NoStop}%
\bibitem [{\citenamefont {Wang}\ and\ \citenamefont
  {Vishwanath}(2006)}]{wang2007}%
  \BibitemOpen
  \bibfield  {author} {\bibinfo {author} {\bibfnamefont {F.}~\bibnamefont
  {Wang}}\ and\ \bibinfo {author} {\bibfnamefont {A.}~\bibnamefont
  {Vishwanath}},\ }\href {\doibase 10.1103/PhysRevB.74.174423} {\bibfield
  {journal} {\bibinfo  {journal} {Phys. Rev. B}\ }\textbf {\bibinfo {volume}
  {74}},\ \bibinfo {pages} {174423} (\bibinfo {year} {2006})}\BibitemShut
  {NoStop}%
\bibitem [{\citenamefont {Liang}\ \emph {et~al.}(1988)\citenamefont {Liang},
  \citenamefont {Doucot},\ and\ \citenamefont {Anderson}}]{Liang1988}%
  \BibitemOpen
  \bibfield  {author} {\bibinfo {author} {\bibfnamefont {S.}~\bibnamefont
  {Liang}}, \bibinfo {author} {\bibfnamefont {B.}~\bibnamefont {Doucot}}, \
  and\ \bibinfo {author} {\bibfnamefont {P.~W.}\ \bibnamefont {Anderson}},\
  }\href {\doibase 10.1103/PhysRevLett.61.365} {\bibfield  {journal} {\bibinfo
  {journal} {Phys. Rev. Lett.}\ }\textbf {\bibinfo {volume} {61}},\ \bibinfo
  {pages} {365} (\bibinfo {year} {1988})}\BibitemShut {NoStop}%
\bibitem [{\citenamefont {Haldane}(1988)}]{haldane1988a}%
  \BibitemOpen
  \bibfield  {author} {\bibinfo {author} {\bibfnamefont {F.~D.~M.}\
  \bibnamefont {Haldane}},\ }\href {\doibase 10.1103/PhysRevLett.61.1029}
  {\bibfield  {journal} {\bibinfo  {journal} {Phys. Rev. Lett.}\ }\textbf
  {\bibinfo {volume} {61}},\ \bibinfo {pages} {1029} (\bibinfo {year}
  {1988})}\BibitemShut {NoStop}%
\bibitem [{\citenamefont {Kaul}\ \emph {et~al.}(2008)\citenamefont {Kaul},
  \citenamefont {Metlitski}, \citenamefont {Sachdev},\ and\ \citenamefont
  {Xu}}]{rkk3}%
  \BibitemOpen
  \bibfield  {author} {\bibinfo {author} {\bibfnamefont {R.~K.}\ \bibnamefont
  {Kaul}}, \bibinfo {author} {\bibfnamefont {M.~A.}\ \bibnamefont {Metlitski}},
  \bibinfo {author} {\bibfnamefont {S.}~\bibnamefont {Sachdev}}, \ and\
  \bibinfo {author} {\bibfnamefont {C.}~\bibnamefont {Xu}},\ }\href {\doibase
  10.1103/PhysRevB.78.045110} {\bibfield  {journal} {\bibinfo  {journal} {Phys.
  Rev. B}\ }\textbf {\bibinfo {volume} {78}},\ \bibinfo {pages} {045110}
  (\bibinfo {year} {2008})}\BibitemShut {NoStop}%
\bibitem [{\citenamefont {Hastings}(2004)}]{Hastings2004}%
  \BibitemOpen
  \bibfield  {author} {\bibinfo {author} {\bibfnamefont {M.~B.}\ \bibnamefont
  {Hastings}},\ }\href {\doibase 10.1103/PhysRevB.69.104431} {\bibfield
  {journal} {\bibinfo  {journal} {Phys. Rev. B}\ }\textbf {\bibinfo {volume}
  {69}},\ \bibinfo {pages} {104431} (\bibinfo {year} {2004})}\BibitemShut
  {NoStop}%
\bibitem [{\citenamefont {Read}\ and\ \citenamefont
  {Sachdev}(1990)}]{Read1990}%
  \BibitemOpen
  \bibfield  {author} {\bibinfo {author} {\bibfnamefont {N.}~\bibnamefont
  {Read}}\ and\ \bibinfo {author} {\bibfnamefont {S.}~\bibnamefont {Sachdev}},\
  }\href {\doibase 10.1103/PhysRevB.42.4568} {\bibfield  {journal} {\bibinfo
  {journal} {Phys. Rev. B}\ }\textbf {\bibinfo {volume} {42}},\ \bibinfo
  {pages} {4568} (\bibinfo {year} {1990})}\BibitemShut {NoStop}%
\bibitem [{\citenamefont {Gros}(1989)}]{GROS1989}%
  \BibitemOpen
  \bibfield  {author} {\bibinfo {author} {\bibfnamefont {C.}~\bibnamefont
  {Gros}},\ }\href {\doibase 10.1016/0003-4916(89)90077-8} {\bibfield
  {journal} {\bibinfo  {journal} {Ann. Phys. (N. Y).}\ }\textbf {\bibinfo
  {volume} {189}},\ \bibinfo {pages} {53} (\bibinfo {year} {1989})}\BibitemShut
  {NoStop}%
\bibitem [{\citenamefont {Tay}\ and\ \citenamefont
  {Motrunich}(2011)}]{Tay2011}%
  \BibitemOpen
  \bibfield  {author} {\bibinfo {author} {\bibfnamefont {T.}~\bibnamefont
  {Tay}}\ and\ \bibinfo {author} {\bibfnamefont {O.~I.}\ \bibnamefont
  {Motrunich}},\ }\href {http://prb.aps.org/abstract/PRB/v84/i2/e020404}
  {\bibfield  {journal} {\bibinfo  {journal} {Phys. Rev. B}\ }\textbf {\bibinfo
  {volume} {84}},\ \bibinfo {pages} {020404(R)} (\bibinfo {year}
  {2011})}\BibitemShut {NoStop}%
\bibitem [{\citenamefont {Sandvik}\ and\ \citenamefont
  {Evertz}(2010)}]{Sandvik2010}%
  \BibitemOpen
  \bibfield  {author} {\bibinfo {author} {\bibfnamefont {A.~W.}\ \bibnamefont
  {Sandvik}}\ and\ \bibinfo {author} {\bibfnamefont {H.~G.}\ \bibnamefont
  {Evertz}},\ }\href {\doibase 10.1103/PhysRevB.82.024407} {\bibfield
  {journal} {\bibinfo  {journal} {Phys. Rev. B}\ }\textbf {\bibinfo {volume}
  {82}},\ \bibinfo {pages} {024407} (\bibinfo {year} {2010})}\BibitemShut
  {NoStop}%
\bibitem [{\citenamefont {Tang}\ \emph {et~al.}(2011)\citenamefont {Tang},
  \citenamefont {Sandvik},\ and\ \citenamefont {Henley}}]{Tang2011}%
  \BibitemOpen
  \bibfield  {author} {\bibinfo {author} {\bibfnamefont {Y.}~\bibnamefont
  {Tang}}, \bibinfo {author} {\bibfnamefont {A.~W.}\ \bibnamefont {Sandvik}}, \
  and\ \bibinfo {author} {\bibfnamefont {C.~L.}\ \bibnamefont {Henley}},\
  }\href {\doibase 10.1103/PhysRevB.84.174427} {\bibfield  {journal} {\bibinfo
  {journal} {Phys. Rev. B}\ }\textbf {\bibinfo {volume} {84}},\ \bibinfo
  {pages} {174427} (\bibinfo {year} {2011})}\BibitemShut {NoStop}%
\bibitem [{\citenamefont {Bishop}\ \emph {et~al.}(2008)\citenamefont {Bishop},
  \citenamefont {Li}, \citenamefont {Darradi},\ and\ \citenamefont
  {Richter}}]{Bishop2008}%
  \BibitemOpen
  \bibfield  {author} {\bibinfo {author} {\bibfnamefont {R.~F.}\ \bibnamefont
  {Bishop}}, \bibinfo {author} {\bibfnamefont {P.~H.~Y.}\ \bibnamefont {Li}},
  \bibinfo {author} {\bibfnamefont {R.}~\bibnamefont {Darradi}}, \ and\
  \bibinfo {author} {\bibfnamefont {J.}~\bibnamefont {Richter}},\ }\href
  {\doibase 10.1088/0953-8984/20/25/255251} {\bibfield  {journal} {\bibinfo
  {journal} {J. Phys. Condens. Matter}\ }\textbf {\bibinfo {volume} {20}},\
  \bibinfo {pages} {255251} (\bibinfo {year} {2008})}\BibitemShut {NoStop}%
\bibitem [{\citenamefont {Yang}\ and\ \citenamefont {Yao}(2012)}]{Yang2012}%
  \BibitemOpen
  \bibfield  {author} {\bibinfo {author} {\bibfnamefont {F.}~\bibnamefont
  {Yang}}\ and\ \bibinfo {author} {\bibfnamefont {H.}~\bibnamefont {Yao}},\
  }\href {\doibase 10.1103/PhysRevLett.109.147209} {\bibfield  {journal}
  {\bibinfo  {journal} {Phys. Rev. Lett.}\ }\textbf {\bibinfo {volume} {109}},\
  \bibinfo {pages} {147209} (\bibinfo {year} {2012})}\BibitemShut {NoStop}%
\bibitem [{\citenamefont {Capponi}\ \emph {et~al.}(2013)\citenamefont
  {Capponi}, \citenamefont {Chandra}, \citenamefont {Auerbach},\ and\
  \citenamefont {Weinstein}}]{Capponi2013}%
  \BibitemOpen
  \bibfield  {author} {\bibinfo {author} {\bibfnamefont {S.}~\bibnamefont
  {Capponi}}, \bibinfo {author} {\bibfnamefont {V.~R.}\ \bibnamefont
  {Chandra}}, \bibinfo {author} {\bibfnamefont {A.}~\bibnamefont {Auerbach}}, \
  and\ \bibinfo {author} {\bibfnamefont {M.}~\bibnamefont {Weinstein}},\ }\href
  {\doibase 10.1103/PhysRevB.87.161118} {\bibfield  {journal} {\bibinfo
  {journal} {Phys. Rev. B}\ }\textbf {\bibinfo {volume} {87}},\ \bibinfo
  {pages} {161118(R)} (\bibinfo {year} {2013})}\BibitemShut {NoStop}%
\bibitem [{\citenamefont {Gong}\ \emph {et~al.}()\citenamefont {Gong},
  \citenamefont {Zhu}, \citenamefont {Sheng}, \citenamefont {Motrunich},\ and\
  \citenamefont {Fisher}}]{Gong2013a}%
  \BibitemOpen
  \bibfield  {author} {\bibinfo {author} {\bibfnamefont {S.-S.}\ \bibnamefont
  {Gong}}, \bibinfo {author} {\bibfnamefont {W.}~\bibnamefont {Zhu}}, \bibinfo
  {author} {\bibfnamefont {D.~N.}\ \bibnamefont {Sheng}}, \bibinfo {author}
  {\bibfnamefont {O.~I.}\ \bibnamefont {Motrunich}}, \ and\ \bibinfo {author}
  {\bibfnamefont {M.~P.~A.}\ \bibnamefont {Fisher}},\ }\href
  {http://arxiv.org/abs/1311.5962} {\ }\Eprint {http://arxiv.org/abs/1311.5962}
  {arXiv:1311.5962} \BibitemShut {NoStop}%
\bibitem [{\citenamefont {Hu}\ \emph {et~al.}(2013)\citenamefont {Hu},
  \citenamefont {Becca}, \citenamefont {Parola},\ and\ \citenamefont
  {Sorella}}]{Hu2013}%
  \BibitemOpen
  \bibfield  {author} {\bibinfo {author} {\bibfnamefont {W.-J.}\ \bibnamefont
  {Hu}}, \bibinfo {author} {\bibfnamefont {F.}~\bibnamefont {Becca}}, \bibinfo
  {author} {\bibfnamefont {A.}~\bibnamefont {Parola}}, \ and\ \bibinfo {author}
  {\bibfnamefont {S.}~\bibnamefont {Sorella}},\ }\href {\doibase
  10.1103/PhysRevB.88.060402} {\bibfield  {journal} {\bibinfo  {journal} {Phys.
  Rev. B}\ }\textbf {\bibinfo {volume} {88}},\ \bibinfo {pages} {060402(R)}
  (\bibinfo {year} {2013})}\BibitemShut {NoStop}%
\end{thebibliography}
%

\end{document}